\documentclass[reprint,pre,showpacs,superscriptaddress]{revtex4-1}
\usepackage{amsmath,amsthm,amssymb,amscd,float,graphicx}
\usepackage[T1]{fontenc}
\usepackage{lmodern}
\usepackage[colorlinks,linkcolor=blue,citecolor=blue,urlcolor=blue]{hyperref}
\newcommand{\al}{\alpha}
\newcommand{\be}{\beta}
\newcommand{\de}{\delta}

\newcommand{\vep}{\varepsilon}
\newcommand{\ga}{\gamma}
\newcommand{\ka}{\kappa}
\newcommand{\la}{\lambda}
\newcommand{\om}{\omega}
\newcommand{\si}{\sigma}

\newcommand{\ze}{\zeta}
\newcommand{\De}{\Delta}
\newcommand{\Ga}{\Gamma}

\newcommand{\RR}{{\mathbb R}}

\newcommand{\cE}{{\mathcal E}}

\newcommand{\pd}{\partial}

\newcommand{\id}{1\hspace{-.25em}{\rm I}}

\newcommand{\ket}[1]{|#1\rangle}
\newcommand{\bra}[1]{\langle#1|}

\newcommand{\mss}{\kern 1pt}

\renewcommand{\le}{\leqslant}
\renewcommand{\ge}{\geqslant}
\newcommand{\tends}[1]{\bbuildrel{\hbox to 2em{\rightarrowfill}}_{#1}^{}}

\newcommand{\tr}{\operatorname{tr}}

\newcommand{\diag}{\operatorname{diag}}

\newcommand{\iu}{\mathrm i}
\newcommand{\e}{\mathrm e}
\newcommand{\diff}{\mathrm{d}}

\newcommand{\f}{\mathrm{f}}
\newcommand{\ord}{\operatorname{o}}
\newcommand{\Or}{\operatorname{O}}
\newcommand{\su}{\mathrm{su}}

\newcommand{\en}{\enspace}
\newcommand{\all}{\forall}
\newcommand{\pdf}[2]{\frac{\partial #1}{\partial #2}}

\newcommand{\Int}[1]{\,\mathop{\!#1}\limits^{\lower1ex\hbox{$\scriptstyle\circ$}}{}}

\theoremstyle{remark}

\begin{document}

\title[Critical behavior of $\su(1|1)$ supersymmetric spin chains with long-range
interactions]{Critical behavior of {\boldmath$\su(1|1)$} supersymmetric spin chains with
  long-range interactions}

\author{Jos\'e A. \surname{Carrasco}}\email{joseacar@ucm.es} \author{Federico
  \surname{Finkel}}\email{ffinkel@ucm.es}
\author{Artemio
  \surname{Gonz\'alez-L\'opez}}\email[Corresponding author. Email address: ]{artemio@ucm.es}
\author{Miguel A. \surname{Rodr\'iguez}}\email{rodrigue@ucm.es} \affiliation{Departamento de
  F\'\i sica Te\'orica II, Universidad Complutense de Madrid, 28040 Madrid, Spain}
\author{Piergiulio
  \surname{Tempesta}}\email{p.tempesta@fis.ucm.es, piergiulio.tempesta@icmat.es}
\affiliation{Departamento de F\'\i sica Te\'orica II, Universidad Complutense de Madrid, 28040 Madrid,
  Spain} \affiliation{Instituto de Ciencias Matem\'aticas \textup(CSIC--UAM--UC3M--UCM\textup), c/ Nicol\'as Cabrera
 13--15, 28049 Madrid, Spain}
\date{March 11, 2016}
\begin{abstract}
  We introduce a general class of $\su(1|1)$ supersymmetric spin chains with long-range
  interactions which includes as particular cases the $\su(1|1)$ Inozemtsev (elliptic) and
  Haldane--Shastry chains, as well as the XX model. We show that this class of models can be
  fermionized with the help of the algebraic properties of the $\su(1|1)$ permutation operator,
  and take advantage of this fact to analyze their quantum criticality when a chemical potential
  term is present in the Hamiltonian. We first study the low energy excitations and the low
  temperature behavior of the free energy, which coincides with that of a $(1+1)$-dimensional
  conformal field theory (CFT) with central charge $c=1$ when the chemical potential lies in the
  critical interval~$(0,\cE(\pi))$, $\cE(p)$ being the dispersion relation. We also analyze the
  von Neumann and R\'enyi ground state entanglement entropies, showing that they exhibit the
  logarithmic scaling with the size of the block of spins characteristic of a one-boson
  $(1+1)$-dimensional CFT. Our results thus show that the models under study are quantum critical
  when the chemical potential belongs to the critical interval, with central charge $c=1$. From
  the analysis of the fermion density at zero temperature, we also conclude that there is a
  quantum phase transition at both ends of the critical interval. This is further confirmed by the
  behavior of the fermion density at finite temperature, which is studied analytically (at low
  temperature), as well as numerically for the~$\su(1|1)$ elliptic chain.
\end{abstract}
\pacs{75.10.Pq, 02.30.Ik, 05.30.-d, 05.30.Rt}
\maketitle

\section{Introduction}
Exactly solvable one-dimensional quantum models are widely used as proving grounds for key ideas
in condensed matter physics and the theory of critical phenomena, since their conceptual
simplicity often makes it possible to derive exact analytical expressions for the relevant
physical quantities. Historically, most of the work in this field has been focused on systems with
short-range interactions, like the well-known Heisenberg and (quantum) Ising chains. In the last
few years, however, it has become feasible to realize in the laboratory quantum spin chains
featuring various types of long-range interactions through different experiments involving, e.g.,
optical lattices of ultracold Rydberg atoms and trapped ions, or neutral atoms in optical
cavities~\cite{PC04,KCIKD09,JLHHZ14,RGLSS14,SZFHC15}. In particular, with the help of hyperfine
``clock'' states of trapped $\vphantom{Yb}^{171}\mathrm{Yb}^+$ ions it is now possible to simulate
quantum spin chains in which the coupling~$h_{ij}$ between the $i$-th and $j$-th sites is
inversely proportional to a power~$\al\in(0,3)$ of their distance~\cite{PC04,RGLSS14}. An
important model of this type is the integrable Haldane--Shastry (HS) chain~\cite{Ha88,Sh88}, whose
sites are the equispaced points~$z_k=\e^{2\pi \iu k/N}$ ($1\le k\le N$) on the unit circle with a
coupling proportional to~$|z_i-z_j|^{-2}$. In fact, this chain is a limiting case of a more
general model due to Inozemtsev, in which the coupling $h_{ij}$ is an elliptic function of the
difference~$i-j$ with real period~$N$~\cite{In90}.

Although the particles in the original~HS chain carried spin~$1/2$, the model was shortly
generalized to~$\su(m)$ spin without losing its remarkable integrability
properties~\cite{HHTBP92}. As a matter of fact, the~$\su(m|n)$ supersymmetric version of the HS
chain, originally introduced by Haldane~\cite{Ha93}, has also been studied in the
literature~\cite{BB06,BBS08}. Of particular interest is the~$\su(m|1)$ HS chain (with $m>1$),
since it is essentially equivalent to an~$\su(m)$ supersymmetric $t$-$J$
model~\cite{Su75,Sc87,BB90} with exchange and transfer energies proportional to~$|z_i-z_j|^{-2}$.
This chain, first introduced by Kuramoto and Yokoyama in the~$\su(2)$ case~\cite{KY91}, is an
exactly solvable model which provides one of the simplest realizations of spin-charge separation.

In this work we introduce a wide class of $\su(1|1)$ supersymmetric spin chains with general
translation-invariant couplings~$h_{ij}>0$ and a chemical potential term. For zero chemical
potential, these models include in particular the supersymmetric elliptic chain studied in
Ref.~\cite{FG14JSTAT} and its two limiting cases, namely the $\su(1|1)$ HS chain and the XX model.
The class of models under study are technically simpler than their $\su(m|1)$ counterparts,
essentially due to the fact that they can be transformed into a system of free spinless fermions
in a straightforward way. However, they still exhibit a sufficiently rich structure which makes it
possible to examine a number of key properties in the theory of quantum critical systems in an
analytic fashion.

More precisely, our main objective is to study whether the models under consideration are quantum
critical for suitable values of the chemical potential, and to determine their corresponding
central charge. As is well known, a characteristic feature of $(1+1)$-dimensional CFTs is the fact
that at low temperature~$T$ their free energy per unit length is approximately given (in
appropriate units) by $f_0-\pi c T^2/(6 v)$, where $f_0$ is a constant and $v$ is the effective
speed of ``sound''~\cite{BCN86,Af86}. Since the low temperature behavior of $f$ is determined by
the low-lying states of the theory, this should also be the case for any one-dimensional quantum
system whose low energy spectrum is described by a $(1+1)$-dimensional CFT. In particular, the
determination of the low temperature behavior of the free energy of a one-dimensional critical
model provides an efficient way of determining the central charge of its underlying CFT. In this
way we have been able to show that if the dispersion relation~$\cE(p)$ is monotonic in the
range~$[0,\pi]$ the models under study are critical when the chemical potential $\la$ belongs to
the open interval~$(0,\cE(\pi))$, with central charge $c=1$. As further confirmation of this
result, we have studied the ground state entanglement entropy, i.e., the entropy of the reduced
density matrix of a block of $L$ consecutive spins when the whole chain is in its ground state.
Indeed, it is well known that in a $(1+1)$-dimensional CFT the R\'enyi and von Neumann entanglement
entropies scale as~$(c/6)(1+1/q)\log L$ and $(c/3)\log L$, respectively, where $c$ is the central
charge and~$q$ is the R\'enyi parameter~\cite{HLW94,CC04JSTAT,CC05}. Thus the entanglement entropy
of a quantum critical one-dimensional system should be proportional to~$\log L$ for $L\gg1$, where
the proportionality constant fixes the central charge of the underlying CFT. Again, we have
verified that when the chemical potential belongs to the open critical interval~$(0,\cE(\pi))$ the
entanglement entropy of the models under consideration scales as that of a $(1+1)$-dimensional CFT
with~$c=1$. We have also examined the behavior of the entanglement entropy and the
zero-temperature fermion density as $\la$ approaches the endpoints of the critical interval,
showing that it is consistent with a quantum phase transition at both ends. For the $\su(1|1)$
chain with elliptic interactions we have studied numerically the fermion density at finite
temperature, finding that its behavior is far more complex when the chemical potential lies in the
critical interval. More precisely, for suitable values of $\la$ inside this interval the fermion
density is not a monotonic function of the temperature, but can rather present up to two extrema.

The paper is organized as follows. In Section~\ref{sec.model} we introduce the class of
supersymmetric spin chains with which this work is concerned, and recall how these models can be
fermionized using the algebraic properties of the~$\su(1|1)$ permutation operator. In
Section~\ref{sec.TDc} we analyze the thermodynamics of the general $\su(1|1)$ chain~\eqref{Hchain}
when the dispersion relation is monotonic in the range $0\le p\le\pi$, showing that at low
temperature it behaves as a $(1+1)$-dimensional CFT with $c=1$. In Section~\ref{sec.gsee} we
outline the computation of the von Neumann and R\'enyi ground state entanglement entropies of the
latter models in terms of the eigenvalues of the ground state correlation matrix.
Section~\ref{sec.asymp} is devoted to deriving asymptotic formulas for these entropies, both when
the size of the block of spins tends to infinity and when the chemical potential approaches the
endpoints of the critical interval. In Section~\ref{sec.FD} we perform a numerical study of the
fermion density of the elliptic~$\su(1|1)$ chain at finite temperature, and determine analytically
its low temperature behavior for arbitrary interactions. Finally, in Section~\ref{sec.sumout} we
summarize our conclusions and discuss some future developments suggested by the present work.

\section{The models}\label{sec.model}
Consider a translation-invariant (closed) spin chain whose $N$ sites are occupied by either a
boson or a (spinless) fermion. If we denote by~$b_i^\dagger$ and~$f_i^\dagger$ the operators that
respectively create a boson or a fermion at the $i$-th site, the Hilbert space of the model is the
$2^N$-dimensional subspace of the infinite-dimensional Fock space determined by the constraints
\begin{equation}\label{rest}
b^\dagger_ib_i+f^\dagger_if_i=1\,,\qquad 1\le i\le N\,.
\end{equation}
We shall take as the model's Hamiltonian the operator~\footnote{Here and it what follows, all sums
  range from~$1$ to~$N$ unless otherwise stated.}
\begin{equation}\label{Hchain}
H=\sum_{i<j}h_N(j-i)(1-S_{ij})-\la N_\f,
\end{equation}
where~$\la\in\RR$, $N_\f=\sum_if^\dagger_if_i$ is the total fermion number operator, $h_N$
is a nonnegative smooth function and~$S_{ij}$ is the~$\su(1|1)$ spin permutation
operator~\cite{Ha93} defined by
\begin{equation*}
  S_{ij}=b^\dagger_ib^\dagger_jb_ib_j+f^\dagger_if^\dagger_jf_if_j+f^\dagger_jb^\dagger_if_ib_j
  +b^\dagger_jf^\dagger_ib_if_j\,.
\end{equation*}
If we denote by~$\ket0$ and~$\ket 1$ respectively the states occupied
by a boson or a fermion, the action of the operator~$S_{ij}$ on the canonical spin basis with
elements~$\ket{s_1}\otimes\cdots\otimes\ket{s_N}\equiv\ket{s_1,\dots,s_N}$, with $s_i\in\{0,1\}$,
is given by
\begin{equation}\label{Sij}
  S_{ij}\ket{\dots,s_i,\dots,s_j,\dots}=(-1)^n\ket{\dots,s_j,\dots,s_i,\dots}\,,
\end{equation}
where~$n=s_i=s_j$ if $s_i=s_j$ while for $s_i\ne s_j$ $n$ equals the number of fermions at the
sites~$i+1,\dots,j-1$. Note that~$S_{ij}$ is invariant under the supersymmetry transformation
$b_i\leftrightarrow f_i$, so that the term~$\sum_{i<j}h_N(i-j)(1-S_{ij})$ in~$H$ is~$\su(1|1)$
supersymmetric, while the last term~$\la N_\f$, i.e, the chemical potential of the fermions,
transforms into~$\la(N-N_{\mathrm f})$ due to the constraints~\eqref{rest}. Furthermore, we shall
exclusively be concerned in this paper with \emph{closed} (i.e., periodic) chains, for
which~$h_N(x)=h_N(N-x)$. It is customary to extend the function~$h_N$ to the whole real line as
an~$N$-periodic function, so that
\begin{equation}\label{hconds}
h_N(x)=h_N(-x)=h_N(x+N)\ge 0\,,\qquad \all x\in\RR\,.
\end{equation}

It was shown in Ref.~\cite{FG14JSTAT} that any chain of the form~\eqref{Hchain} can be recast into
a model of spinless hopping fermions by identifying the boson state~$\ket0$ with the fermion
vacuum. More precisely, we define a new set of fermion creation
operators~$a^\dagger_i=f_i^\dagger b_i$, $1\le i\le N$, which indeed satisfy the canonical
anticommutation relations (CAR) on account of~\eqref{rest}. For instance, we have
\begin{multline*}
  a_i^\dagger a_i+ a_ia_i^\dagger=f_i^\dagger f_ib_ib_i^\dagger+f_i f_i^\dagger b_i^\dagger b_i\\=
  \{f_i^\dagger,f_i\}b_i^\dagger b_i+f_i^\dagger f_i=b_i^\dagger b_i+f_i^\dagger f_i=1\,.
\end{multline*}
The chain sites can now be either empty (i.e., in the state~$\ket 0$) or occupied by a fermion (in
the state~$\ket 1$), and thus the Hilbert space is the whole~$2^N$-dimensional Fock space built
acting on the vacuum~$\ket{0,\dots,0}$ with the operators~$a_i^\dagger$. As first shown by
Haldane~\cite{Ha93}, from Eqs.~\eqref{Sij} and the constraints~\eqref{rest} it follows that
the~$\su(1|1)$ exchange operator~$S_{ij}$ admits the following simple expression in terms of the
new fermion operators~$a_i,a_i^\dagger$:
\begin{equation*}
S_{ij}=1-a^\dagger_ia_i-a^\dagger_ja_j+a^\dagger_ia_j+a^\dagger_ja_i\,.
\end{equation*}
Likewise, 
\begin{equation*}
f^\dagger_if_i=f^\dagger_if_i(b_i^\dagger b_i+f_i^\dagger f_i)=f^\dagger_if_i(b_ib_i^\dagger+f_i^\dagger f_i-1)=
a^\dagger_ia_i
\end{equation*}
(since $f^\dagger_if_i$ is idempotent), so that~$\la N_\f=\la\sum_ia^\dagger_ia_i$ is simply the
chemical potential for the new fermions. Taking into account the latter identities, the
Hamiltonian~\eqref{Hchain} can be rewritten as
\begin{equation}\label{Hf}
H=-\sum_{i,j}h_N(i-j)a^\dagger_ia_j-\la\sum_ia^\dagger_ia_i,
\end{equation}
where we have set~$h_N(0)=-\sum_{j=1}^{N-1}h_N(j)$ (see Ref.~\cite{FG14JSTAT} for more details). This
Hamiltonian describes a system of $N$ hopping (spinless) free fermions on a circle, with hopping
amplitude between the $i$-th and $j$-th sites given by~$h_N(i-j)$ and chemical potential~$\la$. The
translation invariance of this model (encoded in the periodicity of the function~$h$) suggests
introducing the Fourier-transformed operators
\begin{equation}\label{FT}
  c_l=\frac1{\sqrt N}\,\sum_{k=1}^N\e^{-2\pi\iu kl/N}a_k\,,\qquad 0\le l\le N-1\,.
\end{equation}
It can be readily shown that these operators satisfy the CAR, and can therefore be considered as a
new set of fermionic operators; in fact, as we shall see below,~$c^\dagger_l$ creates a fermion
with momentum~$p=2\pi l/N$ ($\bmod\en 2\pi$). It is shown in Ref.~\cite{FG14JSTAT} that~$H$ is
diagonal when written in terms of the new operators~$c_l$ and their adjoints. In fact, we have
\begin{equation}
  \label{Hc}
  H=\sum_{l=0}^{N-1}\Big(\vep_N(l)-\la\Big) c^\dagger_lc_l\,,
\end{equation}
where
\begin{equation}\label{vepl}
  \vep_N(l)=\sum_{j=1}^{N-1}\big[1-\cos(2\pi jl/N)\big]h_N(j)\,.
\end{equation}
Likewise, the system's total momentum operator~$P$ is given
by
\begin{equation*}
  P=\sum_{l=0}^{N-1}\frac{2\pi l}N\,c^\dagger_lc_l\,,
\end{equation*}
which shows that the operator~$c^\dagger_l$ creates a fermion with momentum $2\pi l/N$
($\bmod\en 2\pi$). In this work we shall be concerned with systems for which~$\vep_N(l)$ depends
on $l$ and~$N$ only through the corresponding momentum~$2\pi l/N$, i.e.,
\[
\vep_N(l)=\cE(2\pi l/N)\,,\qquad 0\le l\le N-1\,,
\]
where the \emph{dispersion relation}~$\cE$ is a smooth function defined in the
interval~$[0,2\pi]$. It easily follows from Eq.~\eqref{vepl} that if such a function~$\cE$ exists
it is necessarily unique, and that $\cE(p)=\cE(2\pi-p)$. An important type of interaction~$h_N(x)$
satisfying the above requirement is given by the elliptic function
\begin{equation}
  \label{hell}
  h_N(x)=\bigg(\frac\al\pi\bigg)^2\sinh^2\biggl(\frac\pi\al\biggr)\,\bigg(\wp_N(x)
  -\frac{2\hat\eta_1}{\al^2}\bigg)\,,
\end{equation}
where~$\al>0$ is a real parameter, $\wp_N(x)\equiv\wp(x;N/2,\iu\al/2)$ and
$\hat\eta_1=\zeta(1/2;1/2,\iu N/(2\al))$, $\wp(x;\om_1,\om_3)$ and $\ze(x;\om_1,\om_3)$ denoting
respectively the Weierstrass elliptic and zeta functions with half-periods~$\om_1$ and
$\om_3$~\cite{WW27,OLBC10}. It can be shown~\cite{FG14JSTAT} that the function~\eqref{hell} satisfies the
three conditions in Eq.~\eqref{hconds}. Moreover, since
\begin{equation*}
  \lim_{\al\to0+}h_N(x)=\de_{1,x}+\de_{N-1,x}\,,\quad
  \lim_{\al\to\infty}h_N(x)=\frac{(\pi/N)^2}{\sin^{2}\left(\frac{\pi x}N\right)}\,,
\end{equation*}
the model~\eqref{Hchain} with interaction strength~\eqref{hell} smoothly interpolates between the
Heisenberg (for~$\al=0$) and Haldane--Shastry (for~$\al=\infty$) $\su(1|1)$ chains (with a
chemical potential term added). In fact, the former of these models can be transformed into the
spin~$1/2$ (closed) XX Heisenberg Hamiltonian
\begin{equation*}
  H=\frac 12\sum_{i=1}^N\big(\si_i^x\si_{i+1}^x+\si_i^y\si_{i+1}^y\big)
  +\bigg(1-\frac\la2\bigg)\sum_{i=1}^N(1+\si_i^z)\,,
\end{equation*}
where~$\si_k^a$ is the~$a$-th Pauli matrix acting on the~$k$-th site and
$\si_{N+1}^a\equiv\si_{1}^a$, with the help of the standard Wigner--Jordan
transformation~\cite{Sa11}
\begin{equation*}
  a_k=\si_1^z\cdots\si_{k-1}^z\cdot\frac12(\si_k^x-\iu\si_k^y)\,,\qquad 1\le k\le N\,.
\end{equation*}

The dispersion relation~$\cE(p)$ for the elliptic interaction~\eqref{hell} was computed in closed
form in Ref.~\cite{FG14JSTAT}. More precisely, from Eq.~(2.21b) in the latter reference and the
homogeneity properties of the Weierstrass functions we have
\begin{equation}
  \label{cEp}
  \cE(p)=2\sinh^2(\pi/\al)\bigg[\wp(p)-\bigg(\ze(p)-\frac{\eta_1 p}\pi\bigg)^2
  -\frac{2\eta_1}\pi\bigg],
\end{equation}
where
\begin{equation*}
  \wp(p)\equiv\wp(p;\pi,\iu\pi/\al)\,,\quad \ze(p)\equiv\ze(p;\pi,\iu\pi/\al)\,,
  \quad\eta_1=\ze(\pi)\,.
\end{equation*}
In particular, we see that in this case the dispersion relation is a pure
$2\pi$-periodic~\footnote{It can be easily checked with the help of Legendre's
  identity~$(\eta_1/\al)+\iu\,\zeta(\iu\pi/\al;\pi,\iu\pi/\al)=1/2$ that~$\cE(p)$ is not an
  elliptic function.} function, independent of the number of particles~$N$. Taking the~$\al\to0+$
and~$\al\to\infty$ limits in the above equation for~$\cE(p)$ one recovers the well-known
dispersion relations of the~XX model~\cite{GW95} and the~$\su(1|1)$ Haldane--Shastry chain, namely
\begin{equation}\label{cEXXHS}
  \cE_{\mathrm{XX}}(p)=2(1-\cos p),\qquad \cE_{\mathrm{HS}}(p)=\frac12\,p(2\pi-p)\,.
\end{equation}

\section{Criticality and thermodynamics}\label{sec.TDc}
In this section we shall exploit the equivalence of the~$\su(1|1)$ supersymmetric
chain~\eqref{Hchain} to the free fermion model~\eqref{Hf} to analyze the critical behavior of this
chain as a function of the chemical potential~$\la$. To this end, we first need to determine the
ground state of the model~\eqref{Hf}, which is straightforward from Eq.~\eqref{Hc}. Indeed, it is
obvious from the latter equation that the modes excited in the ground state are precisely those
whose momenta $p=2\pi l/N$ satisfy the condition~$\la>\cE(p)$\,, so that the ground state is
nondegenerate. Strictly speaking, this is only true if we assume that~$\cE(2\pi l/N)\ne\la$ for
$l=0,\dots,N-1$. Indeed, if $\cE(2\pi l/N)=\la$ the mode with momentum $2\pi l/N$ (and
$2\pi(N-l)/N$, if $l>0$ and $l\ne N/2$) can be either present or absent in the ground state, which
is therefore degenerate. Since we shall be exclusively concerned with the thermodynamic
limit~$N\to\infty$, from now on we shall implicitly assume without loss of generality that
$\cE(2\pi l/N)\ne\la$ for $0\le l\le N-1$.

We shall also assume in what follows that the dispersion relation has a positive derivative in the
interval~$(0,\pi)$, so that it is monotonically increasing in the latter interval and reaches its
maximum at~$p=\pi$. This is ``generically'' true, and it certainly holds for the dispersion
relation~\eqref{cEp} of the elliptic interaction~\eqref{hell} ---and, in particular, for the XX
model and the $\su(1|1)$ Haldane--Shastry chains. If this is the case, it is straightforward to
show that \emph{the model is gapless for~$\la\in[0,\cE(\pi)]$}.
\begin{figure}[h]
  \centering
  \includegraphics[width=.8\columnwidth]{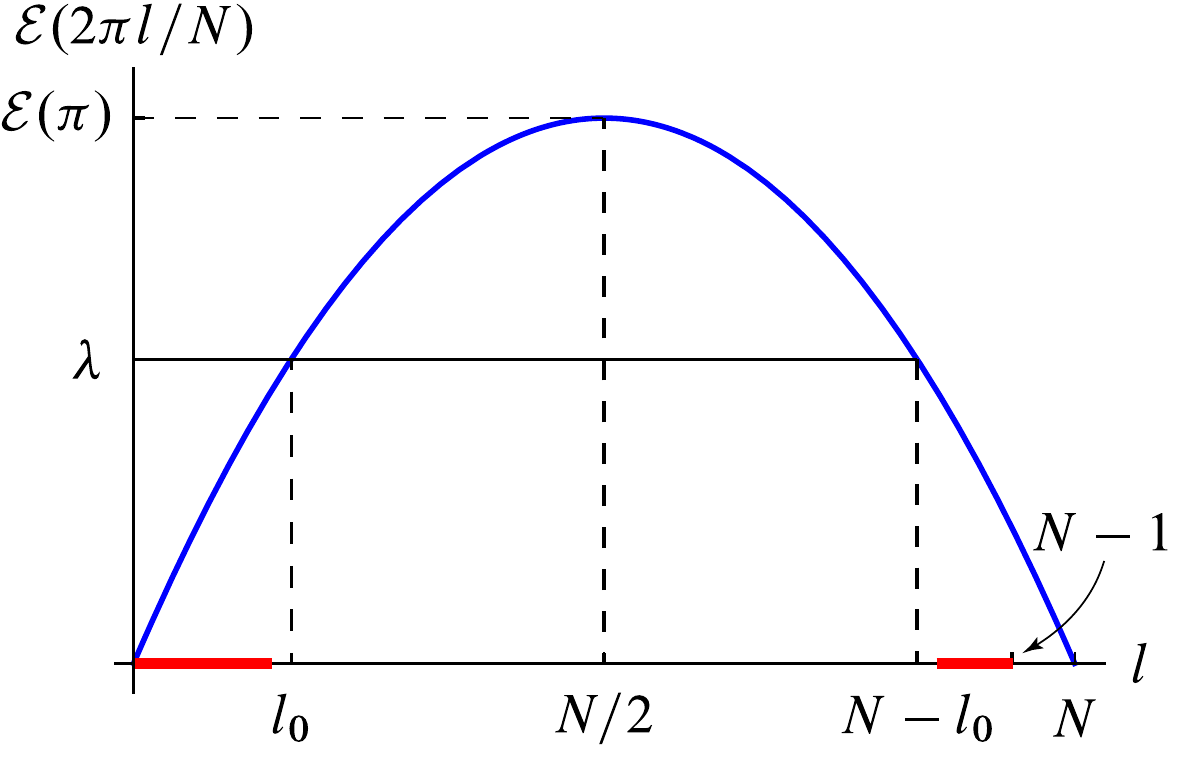}
  \caption{Dispersion relation $\cE(2\pi l/N)$ as a function of the mode number $l=0,\dots,N-1$
    (the range of modes excited in the ground state for the given~$\lambda$ has been represented
    by a thick red line).}
    \label{fig.disp}
  \end{figure}
  
  Indeed, first of all it is clear that the system is gapped for~$\la<0$ or $\la>\cE(\pi)$. For
  instance, for~$\la<0$ the gap between the first excited state~$c_0^\dagger\ket{0,\dots,0}$ and
  the ground state is $\De E=|\la|>0$, which remains positive as $N\to\infty$. Similarly,
  when~$\la>\cE(\pi)$ the gap is approximately equal to~$\De E=\la-\cE(\pi)>0$ independently
  of~$N$. Suppose, on the other hand, that $0\le\la\le\cE(\pi)$, and let $l_0$ be the root of the
  equation~$\cE(2\pi l_0/N)=\la$ in the interval~$[0,N/2]$, which exists and is unique on account
  of the monotonicity of~$\cE$ in the interval~$[0,\pi]$. The modes excited in the ground state
  are now those with $0\le l\le\lfloor l_0\rfloor$ and $N-\lfloor l_0\rfloor\le l\le N-1$,
  where~$\lfloor l_0\rfloor$ denotes the integer part of $l_0$ (cf.~Fig.~\ref{fig.disp}). Thus
  if~$0\le\la\le\cE(\pi)$ the gap between the first excited state and the ground state, given by
\begin{equation*}
  \De E=\min\bigl(\la-\cE(2\pi\lfloor l_0\rfloor/N),\cE(2\pi(\lfloor l_0\rfloor+1)/N)-\la\bigr),
\end{equation*}
is $\operatorname O(1/N)$, since $\la=\cE(2\pi l_0/N)$. Thus $\De E$ tends to zero as~$N\to\infty$
and the system is gapless, as claimed. (In fact, when $l_0$ is an integer the modes with $l=l_0$
or $l=N-l_0$ may or may not be present in the ground state, but this does not affect the ground
state energy and therefore the foregoing argument.)

We shall next show that when the chemical potential~$\la$ belongs to the open interval
$(0,\cE(\pi))$ the $\su(1|1)$ chain~\eqref{Hchain} is indeed critical, or, more precisely, that at
low energies its spectrum is that of a $(1+1)$-dimensional CFT with one free boson. To begin with,
we note that when~$0<\la<\cE(\pi)$ the low energy excitations of
the chain~\eqref{Hchain} are linear in the excitation momentum. Indeed, let
\begin{equation}
  p_0=2\pi l_0/N\equiv\cE^{-1}(\la)\in(0,\pi)\,,
  \label{p0}
\end{equation}
denote the Fermi momentum, where $\cE^{-1}$ is the inverse function of the restriction of the
dispersion relation to the interval~$[0,\pi]$. Adding a fermion with momentum~$p_0+\De p$ (or,
equivalently, $2\pi-p_0-\De p$), with $0<\De p\ll 1$, to the ground state increases the energy by
$\cE(p_0+\De p)-\la=\cE(p_0+\De p)-\cE(p_0)\simeq \cE'(p_0)\De p$. The same excess energy is
approximately obtained when removing from the ground state a fermion with momentum~$p_0-\De p$
(or~$2\pi-p_0+\De p$). Thus for low excitation momenta we have~$\De E\simeq \cE'(p_0)\De p$, as in
a $(1+1)$-dimensional CFT with speed of ``sound''~$v=\cE'(p_0)$.

The simple argument outlined above, based on linearizing the dispersion relation near the Fermi
momentum $p_0$ (or~$2\pi-p_0$) ---the only region in momentum space relevant at low excitation
energies--- does not provide any information on the central charge of the underlying CFT. A more
precise way of establishing the equivalence at low energies of the~$\su(1|1)$ spin
chain~\eqref{Hchain} with a $(1+1)$-dimensional CFT, and in particular of determining its central
charge, is based on the analysis of the chain's free energy. Indeed, as mentioned in the
Introduction, at low temperatures the free energy per unit length of a $(1+1)$-dimensional CFT is
given (in natural units $\hbar=k_{\mathrm B}=1$) by
\begin{equation}\label{fCFT}
f(T)\simeq f_0-\frac{\pi c T^2}{6 v}\,,
\end{equation}
where $f_0=f(0)$ is a constant, $c$ is the central charge and $v$ is the effective speed of sound.
On the other hand, by Eq.~\eqref{Hc} the free energy of the spin chain~\eqref{Hchain} is simply
given by
\[
F(T)=-T\log Z=-T\sum_{l=0}^{N-1}\log Z_l\,,
\]
where~$Z_l=1+\e^{-\be(\cE(2\pi l/N)-\la)}$ (with $\be\equiv1/T$) is the partition function of
the~$l$-th normal mode. Substituting in the previous equation and using the
relation~$\cE(p)=\cE(2\pi-p)$ we obtain the closed formula
\begin{equation}\label{fint}
  f(T)=\lim_{N\to\infty}\frac{F(T)}N=-\frac T\pi\int_0^\pi\log\big[1+\e^{-\be(\cE(p)-\la)}\big]\diff p.
\end{equation}
In order to determine the low temperature behavior of~$f(T)$, we note that~$\cE(p)-\la$ is
negative for~$0<p<p_0$ and positive for~$p_0<p<\pi$, so that $f(T)=f_0+f_1(T)+f_2(T)$, where
\begin{equation}\label{f0def}
f_0=\frac1\pi\int_0^{p_0}[\cE(p)-\la]\,\diff p=f(0)
\end{equation}
is constant
and
\begin{align}
  \label{f1def}
  f_1(T)&=-\frac T\pi\int_0^{p_0}\log\big[1+\e^{-\be(\la-\cE(p))}\big]\diff p\,,\\
  \label{f2def}
f_2(T)&=-\frac T\pi\int_{p_0}^\pi\log\big[1+\e^{-\be(\cE(p)-\la)}\big]\diff p
\end{align}
vanish at~$T=0$. The low temperature behavior of~$f_1(T)$ can be determined by performing the
change of variable $x=(\la-\cE(p))/T$, which yields
\begin{equation}\label{f1x}
f_1(T)=-\frac{T^2}{\pi}\int_0^{\la\be}\log(1+\e^{-x})\frac{\diff x}{\cE'(p)}\,.
\end{equation}
The condition~$\cE'(p_0)\ne0$ implies that~$p-p_0=\Or(Tx)$ and hence $\cE'(p)=v+\Or(Tx)$,
where~$v=\cE'(p_0)$ is the Fermi velocity. We thus have
\[
f_1(T)=-\frac{T^2}{\pi v}\int_0^{\la\be}\log(1+\e^{-x})\,\diff x+\Or(T^3)\,,
\]
so that for~$T\ll 1$ we obtain~\footnote{Indeed, $\log(1+\e^{-x})=\Or(\e^{-x})$ for~$x\to\infty$,
  so that the error incurred in extending the upper integration limit to $+\infty$ is bounded by
  $C\int_{\la\be}^\infty\e^{-x}\diff x=C\e^{-\la\be}\ll T^j$ for all $j>0$, where $C$ is a
  positive constant.}
\begin{align*}
  f_1(T)&=-\frac{T^2}{\pi v}\int_0^{\infty}\log(1+\e^{-x})\,\diff x+\Or(T^3)\\
        &= -\frac{\pi T^2}{12 v}+\Or(T^3)\,.
\end{align*}
The last term~$f_2(T)$ can be similarly dealt with
through the change of variable~$x=\be(\cE(p)-\la)$, with the same result. Hence at low
temperatures we have
\[
f(T)=f_0-\frac{\pi T^2}{6 v}+\Or(T^3)\,,
\]
which coincides with~Eq.~\eqref{fCFT} with~$c=1$. This shows that the spin chain~\eqref{Hchain} is
indeed critical for~$0<\la<\cE(\pi)$, with central charge~$c=1$.

The critical behavior of the~$\su(1|1)$ chain at the endpoints~$\la=0,\cE(\pi)$ can be similarly
investigated. Indeed, suppose to begin with that~$\la=0$. In this case~$f(T)=f_2(T)$, where~$f_2$
is as in Eq.~\eqref{f2def} with~$p_0=0$, so that performing the change of variable~$x=\be\cE(p)$
we obtain
\[
f(T)=-\frac{T^2}{\pi}\int_0^{\be\cE(\pi)}\log(1+\e^{-x})\frac{\diff x}{\cE'(p)}\,.
\]
The dispersion relation can be expanded around $p=0$ as $\cE(p)=(p/a)^{\ka}+\Or(p^{\ka+1})$,
where~$\ka\ge1$ denotes the order of the lowest nonvanishing derivative of~$\cE$ at the origin
(generically, therefore, ~$\ka=1$) and
\begin{equation}\label{adef}
a\equiv\bigg[\frac{\ka!}{\cE^{(\ka)}(0)}\bigg]^{1/\ka}\,.
\end{equation}
From the latter expansion we have~$p/a=(Tx)^{1/\ka}+\Or[(Tx)^{2/\ka}]$, and therefore
\begin{equation}\label{cEpp}
\cE'(p)=\frac\ka a\,\left(\frac pa\right)^{\ka-1}+\Or(p^\ka)=\frac\ka a\,(Tx)^{1-\frac1\ka}+\Or(Tx)\,.
\end{equation}
Substituting into the previous equation for~$f(T)$ we thus obtain
\begin{equation}\label{fTapp}
f(T)= -\frac{a I_\ka}{\ka\pi}\,T^{1+\frac1\ka}+\Or(T^{1+\frac2\ka})\,,\qquad T\ll 1\,,
\end{equation}
with
\[
I_\ka\equiv \int_0^\infty x^{\frac1\ka-1}\,\log(1+\e^{-x})\,\diff x\,.
\]
The integral~$I_\ka$ can actually be evaluated using the technique of Ref.~\cite{FG14JSTAT}, namely:
\begin{align*}
  I_\ka&=\sum_{n=1}^{\infty}\frac{(-1)^{n-1}}n\int_0^\infty x^{\frac1\ka-1}\,\e^{-nx}\diff x\\
     &=
       \Ga(\ka^{-1})\sum_{n=1}^{\infty}\frac{(-1)^{n-1}}{n^{1+\frac1\ka}}\equiv\Ga(\ka^{-1})\eta(1+\ka^{-1})\\
     &=(1-2^{-1/\ka})\Ga(\ka^{-1})\ze_{\mathrm R}(1+\ka^{-1})\,,
\end{align*}
where~$\zeta_{\mathrm R}(z)$ is Riemann's~zeta function, $\eta(z)$ is Dirichlet's eta function, and
we have used the identity~$\eta(z)=(1-2^{1-z})\ze_{\mathrm R}(z)$. Substituting into
Eq.~\eqref{fTapp} we finally obtain
\begin{equation}\label{fTp0}
f(T)=-\ga\,T^{1+\frac1\ka}+\Or(T^{1+\frac2\ka})\,,
\end{equation}
with
\begin{equation}\label{gamma}
\ga=\frac{a}{\pi}\,(1-2^{-1/\ka})\Ga(1+\ka^{-1})\ze_{\mathrm R}(1+\ka^{-1})\,.
\end{equation}
We thus see that for~$\la=0$ the chain~\eqref{Hchain} cannot be critical unless~$\ka=1$, i.e.,
$\cE'(0)\ne0$. Moreover, for~$\ka=1$ we have~$a=1/v$, and therefore
\[
f(T)= -\frac{\pi T^2}{12v}+\Or(T^3)\,.
\]
This shows that when~$\la=0$ and~$\cE'(0)\ne0$ the chain~\eqref{Hchain} is still critical but has
central charge~$c=1/2$, and its low energy behavior is therefore described by a CFT with one free
\emph{fermion}. For instance, for the elliptic interaction~\eqref{hell} $\ka=2$
for~$0\le\al<\infty$, while~$\ka=1$ for~$\al=\infty$. In particular, for~$0\le\al<\infty$
Eqs.~\eqref{fTp0}-\eqref{gamma} with~$\ka=2$ reproduce the result in Ref.~\cite{FG14JSTAT}. On the
other hand, it is well known that the~$\su(1|1)$ Haldane--Shastry chain (i.e., the~$\al=\infty$
case) can indeed be described at low energies by a $(1+1)$-dimensional CFT with one free
fermion~\cite{BBS08}.

The analysis is totally analogous at the other endpoint $\la=\cE(\pi)$. Indeed, since
now~$\cE(p)-\cE(\pi)<0$ for~$0\le p<\pi$ we have $f(T)=f_0+f_1(T)$, where~$f_0$ and~$f_1$ are
respectively given by Eqs.~\eqref{f0def} and~\eqref{f1def} with~$\la=\cE(\pi)$. Performing the
usual change of variable~$x=\be[\cE(\pi)-\cE(p)]$ we thus arrive at Eq.~\eqref{f1x}. Near~$p=\pi$
we have $\cE(\pi)-\cE(p)=[(\pi-p)/b]^\nu+\Or[(\pi-p)^{\nu+1}]$, where $\nu$ denotes the lowest
nonvanishing derivative of the dispersion relation at~$p=\pi$ and
\begin{equation}\label{bdef}
b\equiv\bigg[-\frac{\nu!}{\cE^{(\nu)}(\pi)}\bigg]^{1/\nu}\,.
\end{equation}
Note that, by the symmetry~$\cE(p)=\cE(2\pi-p)$, $\nu$ is necessarily \emph{even}
and~$\cE^{(\nu)}(\pi)<0$. Proceeding as before we readily obtain Eqs.~\eqref{fTp0}-\eqref{gamma}
with $a$ and $\ka$ respectively replaced by $b$ and $\nu$. In particular, since in this
case~$\nu\ge2$ we see that at the endpoint~$\la=\cE(\pi)$ the model~\eqref{Hchain} is not
critical. In summary, our analysis indicates that the latter model is critical
for~$0<\la<\cE(\pi)$, and for~$\la=0$ when~$\cE'(0)\ne0$.

\section{Ground state entanglement entropy}\label{sec.gsee}
We shall study in this section the von Neumann entanglement entropy~$S$ of the ground state of the
$\su(1|1)$ supersymmetric model~\eqref{Hchain}, defined as the von Neumann entropy of the reduced
density matrix~$\rho_L$ of a block of~$L$ consecutive sites when the system is in its ground
state. In other words, if we denote by~$\ket\psi$ the ground state of the chain~\eqref{Hchain}
then~$\rho_L=\tr_{N-L}\ket\psi\bra\psi$, where~$\tr_{N-L}$ denotes the trace over the Hilbert
space of the remaining $N-L$ sites, and the von Neumann entanglement entropy is given by
\begin{equation*}
  S=-\tr(\rho_L\log\rho_L)\,.
\end{equation*}
More generally, we shall also consider the R\'enyi entropy
\begin{equation*}
  S_q=\frac{\log\tr(\rho_L^q)}{1-q}\,,
\end{equation*}
where~$q>0$ is a real parameter, which reduces to that of von Neumann in the $q\to1$ limit. As
pointed out in the Introduction, the von Neumann and R\'enyi ground-state entanglement entropies of
a $(1+1)$-dimensional CFT scale as $r_q\log L$ when $L\to\infty$, where the coefficient $r_q$ is
related to the holomorphic and anti-holomorphic central charges $c$ and $\bar c$ by
$r_q=(1+q^{-1})(c+\bar c)/12$ (with~$q=1$ for the von Neumann entropy)\,. Since the~$\su(1|1)$
supersymmetric chain~\eqref{Hchain} is critical for~$0<\la<\cE(\pi)$, with central charge
$c=\bar c=1$, it is to be expected that for this model
\[
S_q\simeq \frac16\,(1+q^{-1})\log L
\]
in the limit $L\to\infty$. In fact, we shall rigorously establish this asymptotic formula in the
next section (cf.~Eq.~\eqref{Sasymp}).

Before addressing the actual computation of the entanglement entropy of the $\su(1|1)$
chain~\eqref{Hchain}, we note that the result is the same for its ``antiferromagnetic''
version~$-H$. This is most easily proved by considering the equivalent Hamiltonian~\eqref{Hf},
whose ground state entanglement entropy is obviously unchanged if we reverse the roles of the
occupied and empty sites. In other words, the entanglement entropy is the same for the
Hamiltonian~\eqref{Hf} as for its image under the replacement $a_i\leftrightarrow a_i^\dagger$.
Using the CAR and the even character of the interaction~$h$, it is immediate to show that the
latter transformation maps $H$ into $-H-N(\la+h_N(0))$, which establishes our claim.

First of all, it is clear that the ground state is not entangled for $\la$ outside the interval
$[0,\cE(\pi)]$. Indeed, if (for instance) $\la<0$ the ground state is obviously the
vacuum~$\ket{0,\dots,0}$ (i.e., the state with all sites occupied by bosons for the original
Hamiltonian~\eqref{Hchain}), since in this case all the modes have positive
energy~$\cE(2\pi l/N)-\la$. In particular, the ground state is a product state
($\ket 0^{\otimes N}$), and is therefore not entangled. The situation is completely analogous for
$\la>\cE(\pi)$, since in this case $\cE(2\pi l/N)-\la<0$ for all $l$ and therefore all the modes
are excited in the ground state. Thus~$c^\dagger_l\ket\psi=0$ for all $l=0,\dots,N-1$, and
therefore
\begin{equation*}
  a_k^\dagger\ket\psi=\frac1{\sqrt N}\,\sum_{l=0}^{N-1}\e^{-2\pi\iu
    kl/N}c_l^\dagger\ket\psi=0\,,\qquad 1\le k\le N\,.
\end{equation*}
Hence~$\ket\psi=\ket{1,\dots,1}=\ket1^{\otimes N}$ (i.e., the state with all sites occupied by
fermions), which is again a product state and therefore not entangled. (This is also true
when~$\la=\cE(\pi)$ if $N$ is \emph{odd}.) From the previous considerations it follows that the
ground state entanglement entropy of the model~\eqref{Hchain} vanishes for~$\la$ outside the
interval~$[0,\cE(\pi)]$ (and when~$\la=\cE(\pi)$, if $N$ is odd), since in these cases the ground
state is a product state. For this reason, in the rest of this section we shall suppose that~$\la$
belongs to the open critical interval~$(0,\cE(\pi))$.

We shall next find a closed form expression for the entanglement entropy of the $\su(1|1)$
chain~\eqref{Hchain} by applying the method of Ref.~\cite{VLRK03} to the equivalent fermionic
Hamiltonian~\eqref{Hf}. The first step in our computation is the evaluation of the ground-state
correlation matrix~$A$ of the latter model, with matrix elements
\begin{equation*}
  A_{mn}=\bra\psi a^\dagger_m a_n\ket\psi\equiv\langle a^\dagger_m a_n\rangle\,,\qquad 1\le m,n\le N\,.
\end{equation*}
This matrix can be easily determined (in the thermodynamic limit) from the relations
\begin{equation*}
  \langle c^\dagger_jc_k\rangle=
  \begin{cases}
    0\,,& \lfloor l_0\rfloor+1\le j\le N-\lfloor l_0\rfloor-1\\
    \de_{jk}\,,\quad & \text{otherwise,}
  \end{cases}
\end{equation*}
which in turn are a straightforward consequence of the CAR and the conditions
\begin{equation*}
\left\{
\begin{aligned}
    c_j\ket\psi &= 0\,,\qquad \lfloor l_0\rfloor+1\le j\le N-\lfloor l_0\rfloor-1\\
    c^\dagger_j\ket\psi&=0\,,\qquad \text{otherwise}
\end{aligned}
\right.
\end{equation*}
characterizing the ground state. Indeed, from the inverse Fourier transform formula
\begin{equation*}
  a_k=\frac1{\sqrt N}\,\sum_{l=0}^{N-1}\e^{2\pi\iu kl/N}c_l
\end{equation*}
and the previous relations it immediately follows that~\footnote{Although, strictly speaking,
  Eq.~\eqref{Amn} only holds for $m\ne n$, it is obvious that~$A_{mm}$ coincides with the $m\to n$
  limit of the latter equation.}
\begin{align}
  A_{mn}
  &=\frac1N\left(\sum_{l=0}^{\lfloor l_0\rfloor}+\sum_{l=N-\lfloor
    l_0\rfloor}^{N-1}\right)\e^{-2\pi\iu (m-n)l/N}\notag\\
  &=
    \frac1N+\frac2N\sum_{l=1}^{\lfloor l_0\rfloor}\cos\bigl(2\pi(m-n)l/N\bigr)\notag\\
  &\overset{N\gg1\strut}\simeq \frac1\pi\int_0^{p_0}\cos\bigl(p(m-n)\bigr)\,\diff p
    =\frac{\sin\bigl(p_0(m-n)\bigr)}{\pi(m-n)}\,.
    \label{Amn}
\end{align}
 Let us now consider the analogous correlation
matrix~$A_L$ for a block of~$L$ consecutive sites, which by translation invariance we can take as
the first~$L$ ones. By the defining property of the reduced density matrix~$\rho_L$~\cite{NC10},
for $1\le m,n\le L$ we have
\begin{multline*}
  (A_L)_{mn}=\langle a^\dagger_ma_n\rangle_L\equiv\tr_L(a^\dagger_ma_n\rho_L)\\
  =\tr(a^\dagger_ma_n\ket\psi\bra\psi)=\bra\psi a^\dagger_ma_n\ket\psi\equiv A_{mn}\,,
\end{multline*}
where~$\tr_L$ denotes the trace over the Hilbert space of the first~$L$ sites. Thus~$A_L$ is just
the submatrix of~$A$ consisting of its first~$L$ rows and columns. Following Ref.~\cite{VLRK03},
we now consider an alternative basis of fermionic operators whose correlation matrix is diagonal.
More precisely, let~$U=(u_{mn})_{1\le m,n\le L}$ be a unitary matrix diagonalizing the Hermitian
matrix~$A_L$, i.e., satisfying
\begin{equation}\label{UAL}
  UA_LU^\dagger=\diag(\mu_1,\dots,\mu_L)
\end{equation}
where~$\mu_1,\dots,\mu_L\in[0,1]$ are the eigenvalues of~$A_L$. We then define the operators~$g_k$
($1\le k\le L$) by~$g_k=\sum_{m=1}^Lu_{km}^*a_m$; note that~$g_k$, though certainly nonlocal, acts
on the Hilbert space of the first $L$ sites. The operators~$g_k$ and their adjoints satisfy the
CAR by the unitarity of the matrix~$U$, and their correlation matrix is given by
\begin{equation*}
  \langle g_k^\dagger g_l\rangle_L=\mu_k\de_{kl}
\end{equation*}
on account of Eq.~\eqref{UAL}. As shown in Ref.~\cite{JK04}, the latter equation and Wick's
theorem for Gaussian states imply that the correlation matrix factorizes
as~$\rho_L=\otimes_{k=1}^L\varrho_k$, with
\begin{equation*}
  \varrho_k=\mu_kg_k^\dagger g_k+(1-\mu_k)g_kg_k^\dagger\,.
\end{equation*}
The Hilbert space of the system is the tensor product of the two-dimensional spaces spanned by the
vectors~$\ket v_k,g_k^\dagger\ket v_k$ ($1\le k\le N$), where $g_k\ket v_k=0$. Moreover, from the
CAR it easily follows that~$\varrho_k$ is diagonal in the
basis~$\{\ket v_k,g_k^\dagger\ket v_k\}$, with respective eigenvalues~$1-\mu_k$ and~$\mu_k$. Thus
the von Neumann and R\'enyi entropies of~$\varrho_k$ are respectively equal to~$s(\mu_k)$ and
$s_q(\mu_k)$, where
\begin{equation}\label{sr}
  \left\{
\begin{aligned}
  s(x)&=-x\log x-(1-x)\log(1-x)\,,\\
  s_q(x)&=(1-q)^{-1}\log\big[x^q+(1-x)^q\big]\,.
\end{aligned}
\right.
\end{equation}
By the additivity property of both of these entropies we then have
\begin{equation}\label{SRexact}
S=\sum_{k=1}^Ls(\mu_k)\,,\qquad S_q=\sum_{k=1}^Ls_q(\mu_k)\,.
\end{equation}
Equations~\eqref{sr}-\eqref{SRexact}, which are \emph{exact} for any~$L$, make it possible to
evaluate numerically the ground state entanglement entropy of \emph{any}~supersymmetric~$\su(1|1)$
chain of the form~\eqref{Hchain} in \emph{polynomial} time, since they are based on the
diagonalization of the~$L\times L$ matrix with elements~\eqref{Amn}. From the latter equations it also
follows that the entropy of all of these models is a \emph{universal} function of the Fermi
momentum~$p_0$, the difference between two models being manifested only in the dependence of~$p_0$
on the parameter~$\la$ through Eq.~\eqref{p0}.

\section{Asymptotic formulas for the entanglement entropy}\label{sec.asymp}
Equations~\eqref{sr}-\eqref{SRexact} can be used to obtain approximate expressions for the
entanglement entropy of the general~$\su(1|1)$ supersymmetric chain~\eqref{Hchain} in several
interesting regimes. To begin with, we shall investigate the behavior of the entropy as~$\la$
approaches its extreme critical values~$0$ and~$\cE(\pi)$. Suppose, in the first place, that $\la$
tends to zero for fixed~$L$, so that the Fermi momentum~$p_0$ is much smaller than~$1/L$. In this
case all the matrix elements of the correlation matrix~$A_L$ in Eq.~\eqref{Amn} are approximately
equal to~$p_0/\pi$, so that~$A_L=p_0B_L/\pi$, where~$B_L$ is the $L\times L$ matrix with all matrix
elements equal to~$1$. Since the eigenvalues of~$B_L$ are~$0$ (with multiplicity~$L-1$) and ~$L$,
when~$Lp_0\ll1$ the R\'enyi entanglement entropy is approximately given by
\[
S_q\simeq s_q(Lp_0/\pi)\simeq
\begin{cases}
  \dfrac{(Lp_0/\pi)^q}{1-q}\,,&0<q<1\,;\\[4mm]
  \dfrac{q}{q-1}\dfrac{Lp_0}{\pi}\,,\quad& q>1\,.
\end{cases}
\]
For the same reason, when~$Lp_0\ll1$ the von Neumann entropy can be approximated by
\[
S\simeq s(Lp_0/\pi)\simeq-\frac{Lp_0}{\pi}\log\left(\frac{Lp_0}{\pi}\right).
\]
In particular, we see that both~$S_q$ and $S$ are continuous~\footnote{More precisely, since we
  have not actually computed the entanglement entropy for~$\la$ exactly equal to~$0$ the previous
  calculation only shows that the entropy has the same limit as $\la\to0^-$ and~$\la\to0^+$.}
at~$\la=0$. Similarly, suppose now that $p_0$ is close to its upper critical value~$\cE(\pi)$, so
that~$p_0=\pi-\vep$ with~$\vep\ll 1/L$. In this case we have
\[
A_{mn}\simeq-(-1)^{m-n}\frac\vep\pi\,,\qquad m\ne n\,,
\]
while~$A_{nn}=(\pi-\vep)/\pi$. Thus~$A_L=\id-(\vep C_L)/\pi$, where~$C_L$ is the~$L\times L$ matrix
with matrix elements~$C_{mn}=(-1)^{m-n}$. It is easy to check that the eigenvalues of~$C_L$ are
again $0$ (with multiplicity~$L-1$) and~$L$, so that the previous asymptotic expressions for~$S_q$
and~$S$ still hold with~$p_0$ replaced by~$\pi-p_0$.
In particular, this shows that the von Neumann and~R\'enyi entanglement entropies are both
continuous~\footnote{Cf.~the previous footnote.} also at the upper critical value~$\la=\cE(\pi)$.
On the other hand, it is clear that these entropies have a discontinuous first derivative (with
respect to the chemical potential~$\la$) at both endpoints~$\la=0$ and~$\la=\cE(\pi)$. For
instance, for~$0<\la\ll 1$ we have
\begin{equation}\label{cEk0}
p_0\simeq a\,\la^{1/\ka}\,,
\end{equation}
where~$\ka$ is the order of the first non-vanishing derivative of~$\cE$ at~$p=0$ and~$a$ is
defined in~Eq.~\eqref{adef}. Thus $\diff S/\diff\la$ diverges as $\la^{1/\ka-1}|\log\la|$
when~$\la\to0^+$. Similarly, for~$0<q<1$ the derivative of the R\'enyi entropy diverges
as~$\la^{q/\ka-1}$ in this limit, while for~$q>1$ $\diff S_q/\diff\la$ diverges as~$\la^{1/\ka-1}$
for $\ka>1$ and tends to a non-zero finite limit when~$\ka=1$. The situation is similar at the
other endpoint~$\la=\cE(\pi)$, i.e.,
\begin{equation}\label{cEn0pi}
\pi-p_0\simeq b\big(\cE(\pi)-\la\big)^{1/\nu}\,,
\end{equation}
with $b$ defined by Eq.~\eqref{bdef}, except that now~$\nu$ (the order of the lowest nonvanishing
derivative of~$\cE$ at~$p=\pi$) is necessarily even and thus greater than or equal to~$2$. Hence
in all cases the derivatives of~$S$ and $S_q$ diverge as~$\la\to\cE(\pi)^-$. The above analysis
strongly suggests that there is a quantum phase transition at~$\la=0$ and~$\la=\cE(\pi)$ between
an ordered (non-entangled) and a disordered (entangled) ground state, with the entanglement
entropy as the order parameter. This conclusion is confirmed by the behavior of the
zero-temperature fermion~density~$n_{\mathrm f}$, which by translation invariance is simply given
by
\begin{equation}\label{nF}
n_{\mathrm f}=\langle a^\dagger_i a_i\rangle\equiv
A_{ii}=\frac{p_0}\pi
\end{equation}
in the critical interval $0<\la<\cE(\pi)$. Indeed, by Eqs.~\eqref{cEk0}-\eqref{cEn0pi}, near the
two critical points~$\la=0,\cE(\pi)$ the fermion density respectively behaves as
$(a/\pi)\la^{1/\ka}$ and $1-(b/\pi)(\cE(\pi)-\la)^{1/\nu}$. Since~$n_{\mathrm f}=0$ for $\la<0$
and $n_{\mathrm f}=1$ for $\la>\cE(\pi)$, this behavior is typical of a quantum phase transition
with exact exponents~$1/\ka$ and~$1/\nu$ at the critical points~$\la=0$ and~$\la=\cE(\pi)$. For
instance, for the elliptic interaction~\eqref{hell} it is known~\cite{FG14JSTAT} that~$\nu=2$ and
$\ka=2$ for~$0\le\al<\infty$, while~$\ka=1$ for~$\al=\infty$ (i.e., for the $\su(1|1)$ HS chain).
The parameters~$a$ and~$b$ can also be exactly computed in this case, namely
\begin{align*}
  a&=\frac{\pi}{\sinh(\pi/\al)}\,\left(\frac{\pi^2}{6}g_2-2\eta_1^2\right)^{-1/2}\,,
  \\
  b&=\frac{\pi}{\sinh(\pi/\al)}\,\left[\pi^2\bigg(\frac{g_2}{2}-4e_1^2\bigg)+2\eta_1(\eta_1+2\pi e_1)
     \right]^{-1/2},
\end{align*}
where~$e_1=\wp(\pi)$ and~$g_2$ is the second invariant of the Weierstrass function with
half-periods~$(\pi,\iu\,\pi/\al)$~\cite{OLBC10}.
\begin{figure}[h]
  \centering
  \includegraphics[width=.9\columnwidth]{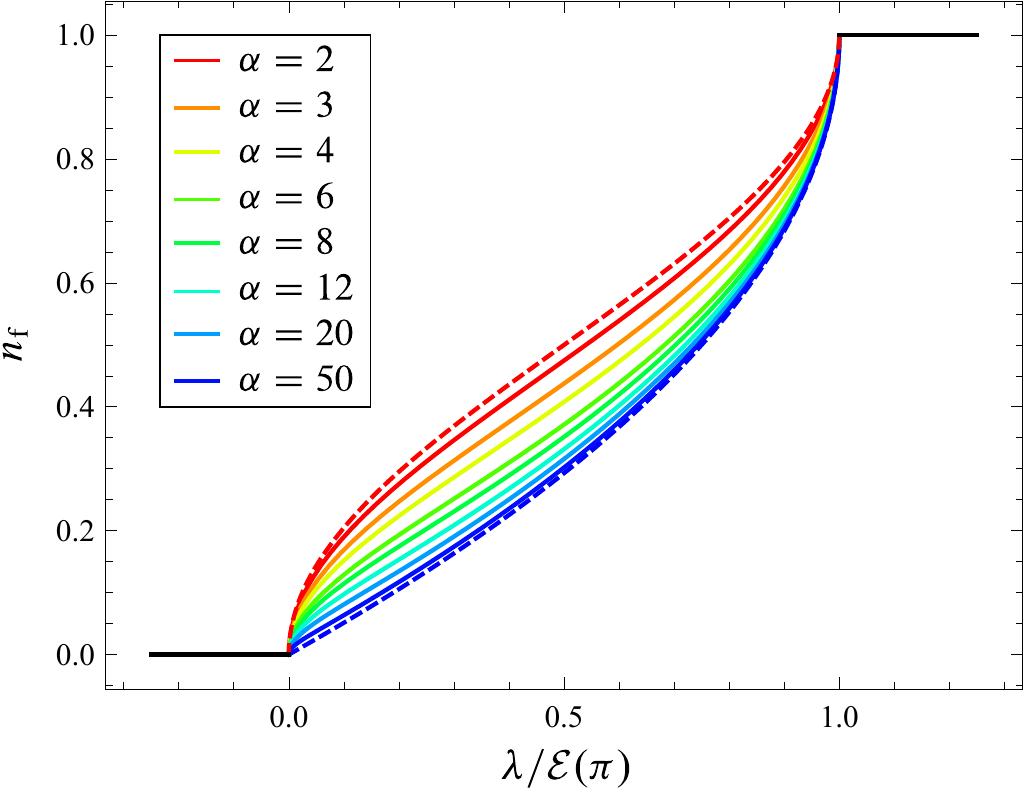}
  \caption{Zero temperature density of fermions of the $\su(1|1)$ chain~\eqref{Hchain} with
    elliptic interactions~\eqref{hell} for several values of the parameter~$\al$ in the range
    $[2,50]$. The red (top) and blue (bottom) dashed curves correspond respectively to the XX
    model ($\al=0$) and the $\su(1|1)$ Haldane--Shastry chain ($\al=\infty$).}
  \label{fig.nF}
\end{figure}
For the general elliptic~$\su(1|1)$ model with interactions~\eqref{hell} (with $0<\al<\infty$) and
dispersion relation~\eqref{cEp}, it is of course unfeasible to explicitly invert~$\cE$ to obtain a
closed-form expression for the Fermi momentum~$p_0=\cE^{-1}(\la)$. Note, however, that the graph
of the fermion density $n_{\mathrm f}$ admits the simple parametrization $(\cE(p),p/\pi)$, with
$0<p<\pi$. In this way we have generated the plot in Fig.~\ref{fig.nF}, where~$n_{\mathrm f}$ is
represented as a function of the normalized parameter~$\la/\cE(\pi)$, where
$\cE(\pi)=2\sinh^2(\pi/\al)[e_1-(2\eta_1/\pi)]$, for several values of~$\al$ in the range~$[0,50]$
and for~$\al=\infty$. The fermion density can be easily computed in closed form for the limiting
cases~$\al=0$ and~$\al=\infty$, i.e., for the~XX model and the~$\su(1|1)$~Haldane--Shastry chain,
due to the simple form of their dispersion relations. Indeed, from Eq.~\eqref{cEXXHS} we
immediately obtain
\[
n_{\mathrm f,\mathrm{XX}}=\frac2\pi\arcsin(\sqrt\la/2)\,,\qquad
n_{\mathrm f,\mathrm{HS}}=1-\sqrt{1-\frac{2\la}{\pi^2}}
\]
respectively for $0<\la<4$ and $0<\la<\pi^2/2$. As expected, the first of these formulas agrees
with the result in Ref.~\cite{JK04}, taking into account that our parameter~$\la$ is related to
the parameter~$h$ in the latter reference by $h=2-\la$. On the other hand, the formula for
the~$\su(1|1)$ HS chain is, to the best of our knowledge, new.

It is also of interest to determine the asymptotic behavior of the von Neumann and~R\'enyi entropies
for~$0<\la<\cE(\pi)$ fixed and $L\gg1$. To this end, we note that Eq.~\eqref{Amn} implies that
$A_{mn}$ is a function of~$m-n$ only, and hence the correlation matrix~$A_L$ is a Toeplitz matrix.
This fact can be exploited to find a simple asymptotic formula for the von Neumann and R\'enyi
entanglement entropies in the~$L\to\infty$ limit, as shown in Ref.~\cite{JK04} for the XX model.
The formula in the latter reference, which is based on a particular case of the general
Fisher--Hartwig conjecture~\cite{FH68} proved by E. Basor in 1979~\cite{Ba79}, is also valid for a
general model of the form~\eqref{Hchain} provided only that we express the result in terms of the
Fermi momentum~$p_0$. Indeed, this formula relies only on Eq.~\eqref{Amn} for the correlation
matrix which, as we have just seen, holds for the model~\eqref{Hchain} with $p_0=\cE^{-1}(\la)$.
In this way one obtains the following asymptotic formula for the R\'enyi entropy in the limit
$L\sin p_0\gg1$:
\begin{equation}
  S_q=\frac{q+1}{6q}\log(L\sin p_0)+\ga_1^{(q)}+\ord(1)\,,
\label{Sasymp}
\end{equation}
while the corresponding formula for the von Neumann entropy is obtained from the above by setting
$q=1$. Here $\ord(1)$ denotes a function of $L$ and $p_0$ which tends to $0$ as
$L\sin p_0\to\infty$, and $\ga_1^{(q)}$ is a constant (independent of $L$ and $p_0$) whose precise
value, which can be found in Ref.~\cite{JK04}, will not be needed in what follows.

Equation~\eqref{Sasymp} can be easily applied in the case of the~XX
and~$\su(1|1)$~Haldane--Shastry chains. Indeed, for the former of these models we have
$\sin p_0=\sqrt{\la(1-\frac\la4)}$, so that~\eqref{Sasymp} agrees with the result in
Ref.~\cite{JK04}. On the other hand, for the $\su(1|1)$ HS chain
$\sin p_0= \sin(\sqrt{\pi^2-2\la}\,)$, and hence Eq.~\eqref{Sasymp} yields the following
asymptotic formulas for the von Neumann and R\'enyi ground state entanglement entropies:
\begin{equation}
  S_q=\frac{q+1}{6q}\log\Bigl[L\sin\Bigl(\sqrt{\pi^2-2\la}\,\Bigr)\Bigr]
  +\ga_1^{(q)}+\ord(1)\,.
\label{SHS}
\end{equation}
These formulas are valid for~$\la$ belonging to the critical interval~$(0,\pi^2/2)$, in the
asymptotic regime $L\sin\sqrt{\pi^2-2\la}\gg 1$. For the general elliptic~$\su(1|1)$
model~\eqref{hell} with $0<\al<\infty$ no such closed formulas are available. However, as for the
fermion density, the graph of~$S_q$ admits the simple parametrization
\begin{equation*}
  \bigg(\cE(p),\frac{q+1}{6q}\log(L\sin p)+\ga_1^{(q)}\bigg),\qquad 0<p<\cE(\pi)\,,
\end{equation*}
where for simplicity we have dropped the~$\ord(1)$ terms.
\begin{figure}[h]
  \centering
  \includegraphics[width=\columnwidth]{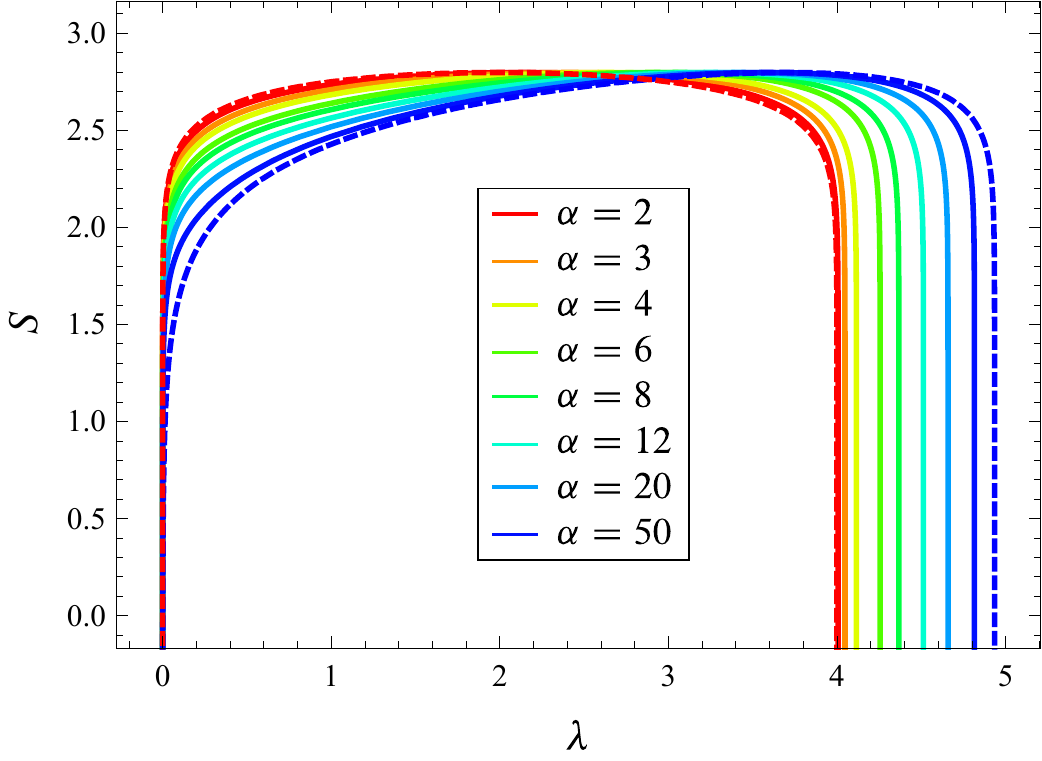}
  \caption{Approximation~\eqref{Sasymp} to the von Neumann entanglement entropy ($q=1$) of the
    elliptic~$\su(1|1)$ chain~\eqref{Hchain}-\eqref{hell} for $L=1000$ and several values of the
    parameter~$\al$ between $2$ and $50$. The red and blue dashed curves correspond respectively
    to the XX Heisenberg model ($\al=0$) and the $\su(1|1)$ Haldane--Shastry chain
    ($\al=\infty$).}
  \label{fig.Splot}
\end{figure}
For instance, in Fig.~\ref{fig.Splot} we present a plot of the approximation~\eqref{Sasymp} to the
von Neumann entropy of the elliptic~$\su(1|1)$ chain~\eqref{Hchain}-\eqref{hell} for $L=1000$ and
several values of the parameter~$\al$, including the limiting cases $\al=0$ (the Heisenberg XX
model) and $\al=\infty$ (the $\su(1|1)$ Haldane--Shastry chain). It is apparent that all of these
plots are qualitatively similar, although only in the case of the XX model ($\al=0$) the graph of
$S$ is symmetric about the midpoint $\la=\cE(\pi)/2$. More precisely, the maximum of $S$ at
$\la=\cE(\pi/2)$ is increasingly displaced towards the right as~$\al$ tends to infinity,
with~$\cE(\pi/2)/\cE(\pi)$ varying continuously from $1/2$ to~$3/4$ as~$\al$ ranges from~$0$
to~$\infty$.

\section{Fermion density at finite temperature}\label{sec.FD}
In the previous sections we have seen that the $\su(1|1)$ chain~\eqref{Hchain} is critical
for~$0<\la<\cE(\pi)$, with central charge~$c=1$. This is confirmed by the asymptotic behavior of
the ground state entanglement entropy when the size of the block of spins considered tends to
infinity. In this section we shall show that the fermion density at finite temperature, given by
\begin{align}
  n_{\mathrm f}&=\lim_{N\to\infty}\frac1N\sum_{l=0}^{N-1}\left(1+\e^{\be(\cE(2\pi l/N)-\la)}\right)^{-1}\notag\\
               &=\frac1\pi\int_0^\pi\frac{\diff p}{1+\e^{\be(\cE(p)-\la)}}\,,
               \label{nfexact}              
\end{align}
also exhibits a qualitatively richer behavior when~$\la$ lies in the critical
interval~$(0,\cE(\pi))$.

\begin{figure}[h]
\centering
  \includegraphics[width=.8\columnwidth]{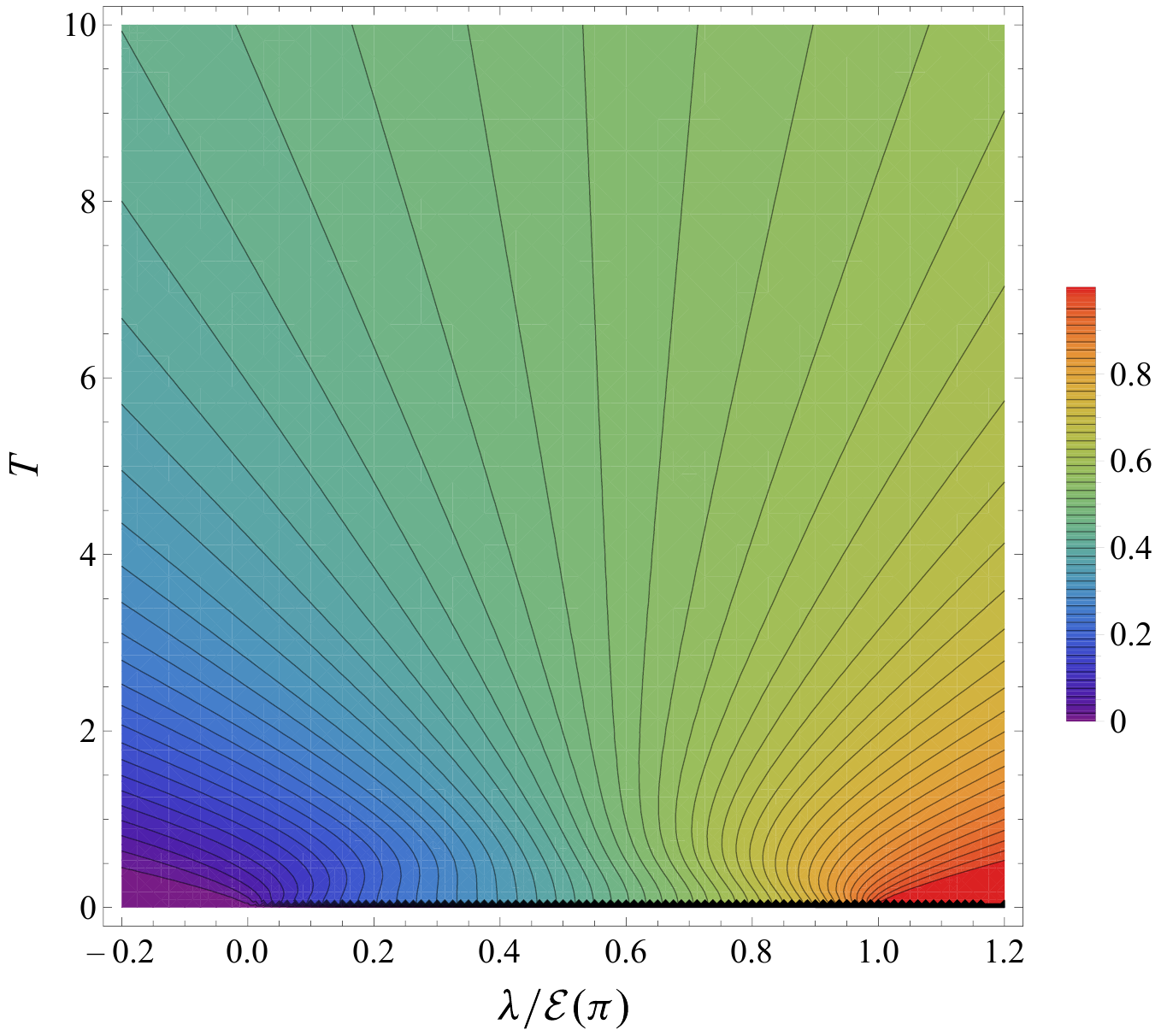}
  \caption{Contour plot of the fermion density of the elliptic $\su(1|1)$
    chain~\eqref{Hchain}-\eqref{hell} with~$\al=5$ for $0\le T\le 10$
    and~$-0.2\le\la/\cE(\pi)\le 1.2$.}
  \label{fig.cplot}
\end{figure}
\begin{figure}[h]
  % \centering
  \includegraphics[width=.75\columnwidth,height=.8\columnwidth]{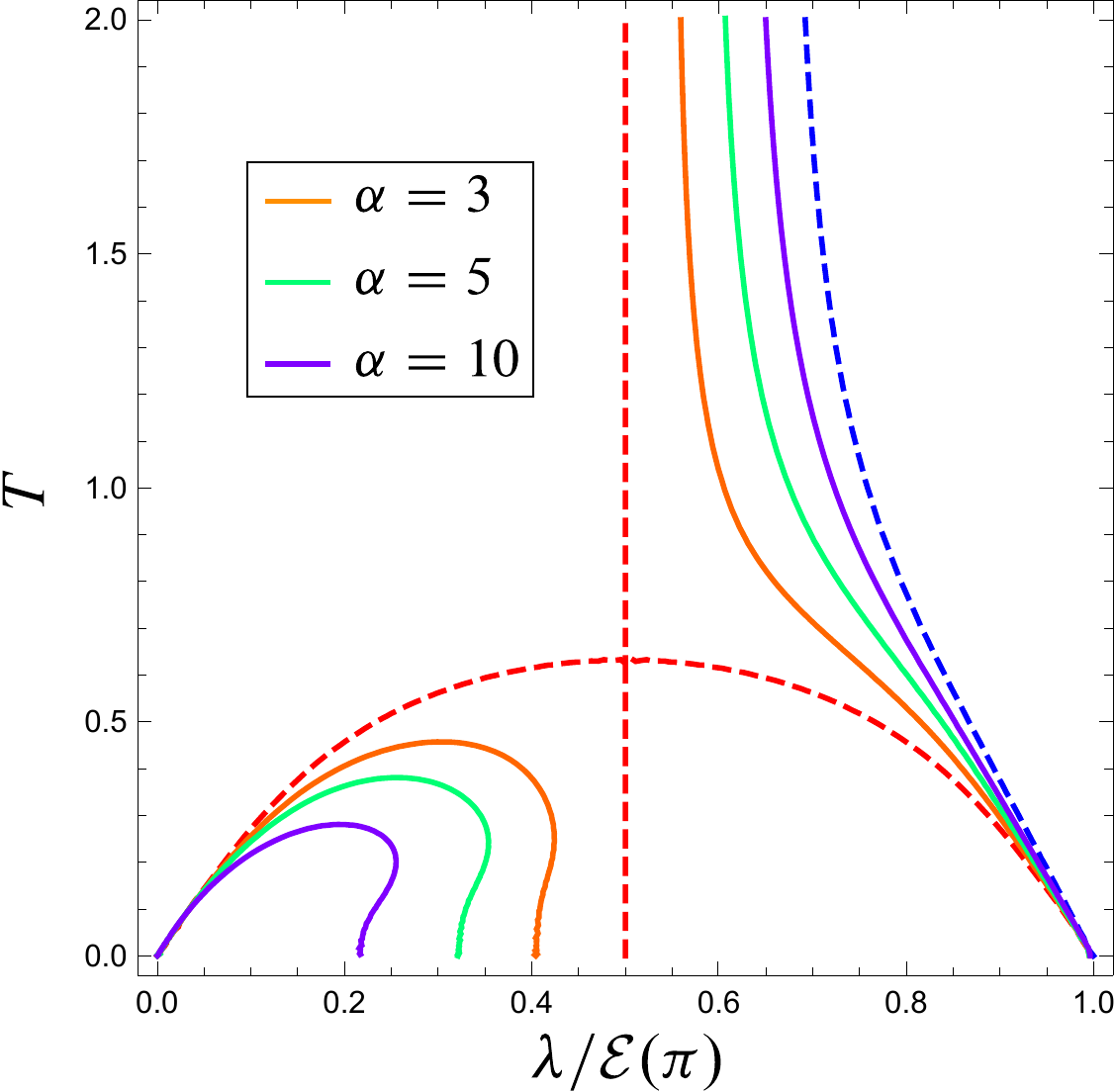}
  \caption{Plot of the implicit curve~$\pd n_{\mathrm f}/\pd T=0$ for the elliptic
    chain~\eqref{Hchain}-\eqref{hell} and several values of the parameter~$\alpha$. The red and
    blue dashed curves correspond respectively to the XX model ($\al=0$) and the
    $\su(1|1)$ Haldane--Shastry chain ($\al=\infty$).}
  \label{fig.dnfplot}
\end{figure}

\begin{figure}[h]
  % \centering
  \includegraphics[width=.9\columnwidth]{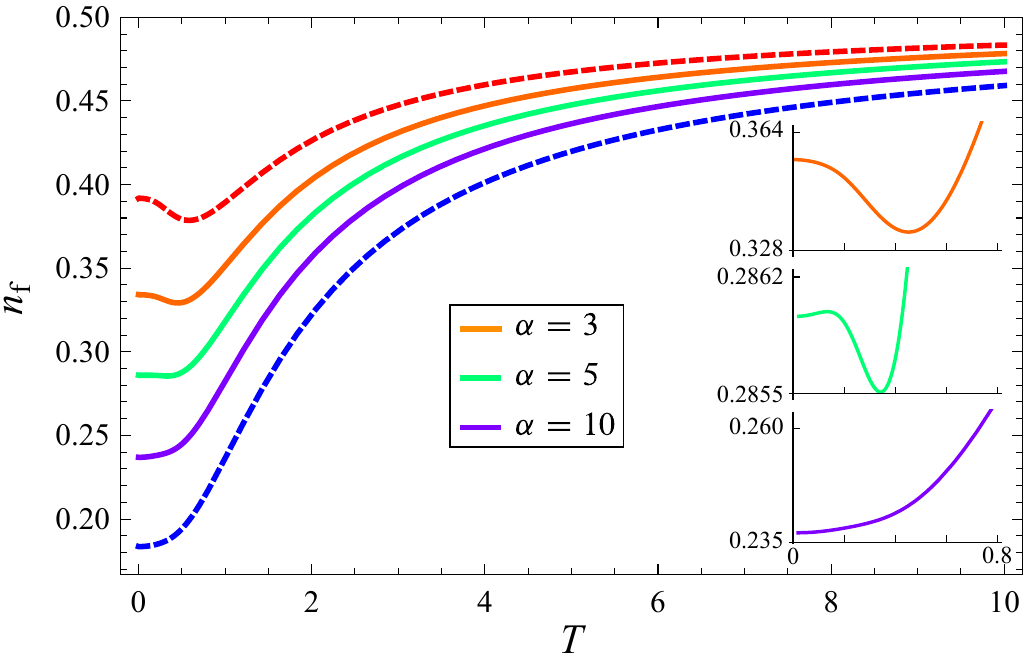}\\[1mm]
  \includegraphics[width=.9\columnwidth]{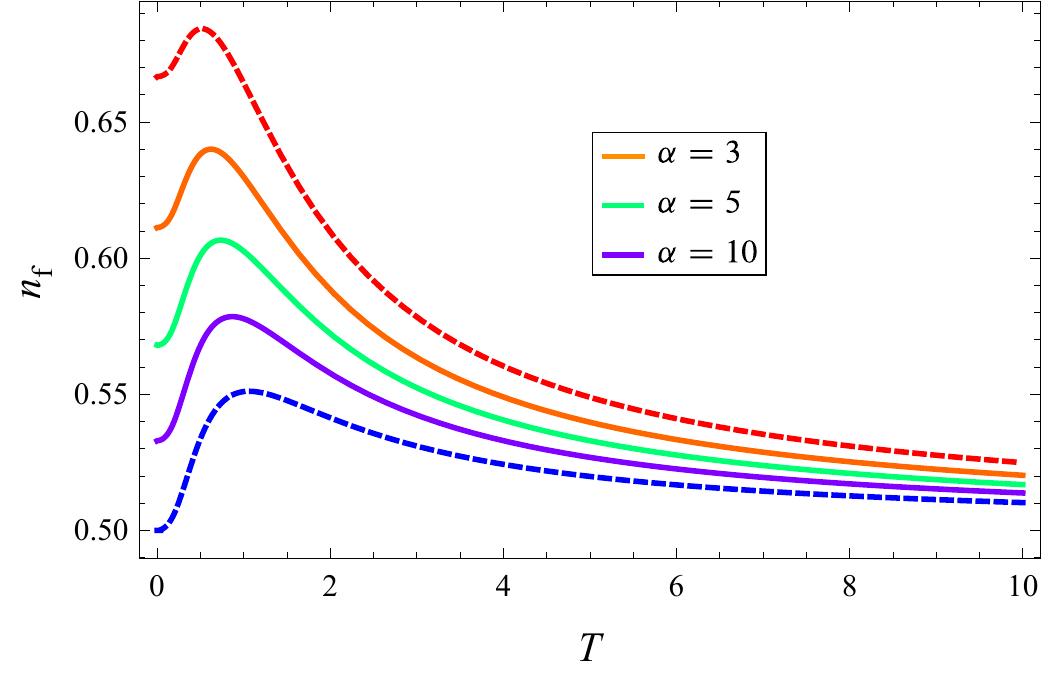}
  \caption{Top: plot of the fermion density of the chain~\eqref{Hchain}-\eqref{hell} as a function
    of the temperature~$T$ for several values of the parameter~$\al$ and~$\la=\cE(\pi)/3$ (inset:
    blowup of the range~$0\le T\le0.8$). Bottom: analogous plot for~$\la=3\cE(\pi)/4$. In both
    plots, the red and blue dashed lines correspond to the limiting cases $\al=0$
    and~$\al=\infty$.}
  \label{fig.nfplots}
\end{figure}

%
As a concrete example, we shall first focus on the~$\su(1|1)$ elliptic
chain~\eqref{Hchain}-\eqref{hell}. In Fig.~\ref{fig.cplot} we present a contour plot
of~$n_{\mathrm f}$ for this model with~$\al=5$ for~$T\in[0,10]$ and~$\la/\cE(\pi)\in[-0.2,1.2]$,
obtained by numerically evaluating the integral in Eq.~\eqref{nfexact}. For~$\la$ outside the
critical interval~$(0,\cE(\pi))$, it is clear that $n_{\mathrm f}$ is a monotonic function of~$T$
(increasing for~$\la\le0$, decreasing for~$\la\ge\cE(\pi)$), since
\[
\pdf{n_{\mathrm f}}{T}
=\frac{\be^2}{4\pi}\int_0^{\pi}\frac{\cE(p)-\la}{\cosh^2\bigl[\be(\cE(p)-\la)/2\bigr]}\,\diff p\,.
\]
On the other hand, it is apparent from Fig.~\ref{fig.cplot} that there is a range of values
of~$\la$ in the interval~$(0,\cE(\pi))$ for which the fermion density is not a monotonic function
of the temperature. Remarkably, the $\su(1|1)$ elliptic chain exhibits this interesting behavior
for all values of the parameter~$\al$, including the limiting cases $\al=0$ and~$\al=\infty$. More
precisely, for each $\al$ there are three critical values~$\la_i$ ($i=1,2,3$) of the chemical
potential~$\la$ such that: i)~for~$0<\la\le\la_1$, the fermion density reaches an absolute minimum
at some positive temperature and then increases monotonically towards its limiting value~$1/2$;
ii)~for~$\la_1<\la<\la_2$, $n_{\mathrm f}$ first reaches a maximum at some $T>0$ and then a
minimum, after which it tends monotonically to~$1/2$; iii)~for $\la_2\le\la\le\la_3$,
$n_{\mathrm f}$ is monotonically increasing, and iv)~for $\la_3<\la<\cE(\pi)$, the fermion density
attains an absolute maximum at some $T>0$ and then decreases monotonically towards $1/2$. This is
also true for the limiting values~$\al=0$ (XX model) and $\al=\infty$ ($\su(1|1)$ HS chain), for
which $\la_1=\la_2=\la_3=\cE(\pi)/2=2$ and $\la_1=\la_2=0$, $\la_3=2\cE(\pi)/3=\pi^2/3$,
respectively. This behavior is qualitatively apparent from Fig.~\ref{fig.dnfplot}, where we have
represented the implicit curve~$\pd n_{\mathrm f}/\pd T=0$ vs.~$\la$ and $T$ for
$\al=0,3,5,10,\infty$, and is also confirmed by the plots of $n_{\mathrm f}$ vs.~$T$ for these
values of~$\al$ and $\la=\cE(\pi)/3,3\cE(\pi)/4$ presented in Fig.~\ref{fig.nfplots}.

Although in general the integral in Eq.~\eqref{nfexact} cannot be computed in closed form, its low
temperature behavior can be readily determined, as we shall now explain. To begin with,
when~$\la<0$ the exponent~$\be(\cE(p)-\la)$ is positive throughout the whole integration range, so
that
\[
n_{\mathrm f}\simeq\frac1\pi\int_0^\pi\e^{-\be(\cE(p)-\la)}\diff p=\frac{T\e^{-|\la|\be}}\pi
\int_0^{\be\cE(\pi)}\frac{\e^{-x}}{\cE'(p)}\,\diff x\,,
\]
where~$x=\be\cE(p)$. Using Eq.~\eqref{cEpp} and extending the integration range to~$+\infty$ we
obtain
\begin{multline*}
  n_{\mathrm
    f}\simeq\frac{a}{\ka\pi}\,T^{1/\ka}\e^{-|\la|\be}\int_0^{\infty}x^{\frac1\ka-1}\e^{-x}\,\diff
  x\\= \frac{a}{\pi}\,\Ga(1+\ka^{-1})\,T^{1/\ka}\e^{-|\la|\be}\,,\qquad \la<0\,.
\end{multline*}
Proceeding in a similar way we obtain an analogous formula when~$\la>\cE(\pi)$:
\[
n_{\mathrm f}\simeq1-\frac{b}{\pi}\,\Ga(1+\nu^{-1})\,T^{1/\nu}\e^{-\be(\la-\cE(\pi))}\,,\qquad
\la>\cE(\pi)\,.
\]
We thus see that for $\la\notin[0,\cE(\pi)]$ the fermion density at low temperature is monotonic,
approaching exponentially its zero temperature values $0$ (for $\la<0$) and~$1$
(for~$\la>\cE(\pi)$). For~$\la=0$, the change of variable~$x=\be\cE(p)$ and Eq.~\eqref{cEpp}
easily yield
\begin{align*}
  n_{\mathrm f}
  &\simeq \frac{aT^{1/\ka}}{\ka\pi}\int_0^\infty\frac{x^{\frac1\ka-1}\diff x}{1+\e^x}
    =\frac{aT^{1/\ka}}{\ka\pi}\int_0^\infty\frac{x^{\frac1\ka-1}\e^{-x}}{1+\e^{-x}}\,\diff x\\
  &=\frac{aT^{1/\ka}}{\ka\pi}\sum_{n=1}^\infty(-1)^{n+1}\int_0^\infty x^{\frac1\ka-1}\e^{-nx}\diff x\\
  &=\frac{a}{\pi}\,\Ga(1+\ka^{-1})\eta(\ka^{-1})\,T^{1/\ka}\,,
\end{align*}
and therefore
\[
 n_{\mathrm f}=
 \begin{cases}
   \frac{\log 2}{\pi v}\,T\,,& \ka=1\,,\\[1mm]
   \frac{a}{\pi}\,(1-2^{1-1/\ka})\Ga(1+\ka^{-1})\ze_{\mathrm R}(\ka^{-1})\,T^{1/\ka}\,,&\ka>1\,.
 \end{cases}
\]
Likewise, at the other endpoint~$\la=\cE(\pi)$ we have
\[
n_{\mathrm f}=
1-\frac{b}{\pi}\,(1-2^{1-1/\nu})\Ga(1+\nu^{-1})\ze_{\mathrm R}(\nu^{-1})\,T^{1/\nu}\,,
\]
since now~$\nu$ is even and hence greater than $1$.

Suppose next that $\la$ lies in the critical interval~$(0,\cE(\pi))$. We start by writing the
fermion density as the sum
\begin{align*}
  n_{\mathrm f}&=\frac{p_0}\pi
  -\frac1\pi\int_0^{p_0}\frac{\diff p}{1+\e^{-\be(\cE(p)-\la)}}
                 +\frac1\pi\int_{p_0}^\pi\frac{\diff p}{1+\e^{\be(\cE(p)-\la)}}\\
  &\equiv\frac{p_0}\pi-n_{\mathrm f,1}+n_{\mathrm f,2}\,,
\end{align*}
where the first term is the value of~$n_{\mathrm f}$ at~$T=0$ (cf.~Eq.~\eqref{nF}). In this case
the leading ($\Or(T)$) contributions to the two integrals $n_{\mathrm f,i}$ cancel each other, so
that we need to evaluate the $\Or(T^2)$ corrections. For the first integral, after performing the
change of variable $x=\be(\la-\cE(p))$ we have
\[
n_{\mathrm f,1}=\frac T\pi\int_0^{\la\be}\frac1{\cE'(p)}\,\frac{\diff x}{1+\e^{x}}\,.
\]
Expanding~$\cE'(p)$ to first order in~$p-p_0$ we obtain
\begin{align*}
  \cE'(p)&=\cE'(p_0)+\cE''(p_0)(p-p_0)+\Or[(p-p_0)^2]\\
  &=v-\frac{\cE''(p_0)}v\,Tx+\Or[(Tx)^2]\,,
\end{align*}
where we have used the expansion~$Tx=\cE(p_0)-\cE(p)=-v(p-p_0)+\Or[(p-p_0)^2]$. We thus have
\[
\cE'(p)^{-1}=\frac1v\,\left(1+\frac{\cE''(p_0)}{v^2}\,Tx+\Or[(Tx)^2]\right).
\]
Substituting in the definition of~$n_{\mathrm f,1}$ and using the estimate
\[
\int_{\la\be}^\infty x^j(1+\e^x)^{-1}\diff x\le \int_{\la\be}^\infty
x^j\e^{-x}\diff x=\Or(\be^j\e^{-\la\be})
\]
(with $j=0,1,\dots$) we obtain
\begin{align}
  n_{\mathrm f,1}&=\frac T{\pi v}\int_0^{\infty}\frac{\diff x}{1+\e^{x}}
  +\frac{\cE''(p_0)T^2}{\pi v^3}\int_0^{\infty}\frac{x\,\diff x}{1+\e^{x}}
                   +\Or(T^3)\notag\\
                 &=\frac{\log 2}{\pi v}\,T+\frac{\pi\cE''(p_0)}{12v^3}\,T^2+\Or(T^3)\,.
                   \label{nf1}
\end{align}
The term~$n_{\mathrm f,2}$ can be similarly dealt with through the analogous change of
variable~$x=\be(\cE(p)-\la)$, so that $Tx=\cE(p)-\cE(p_0)=v(p-p_0)+\Or[(p-p_0)^2]$ and hence
\[
\cE'(p)^{-1}=\frac1v\,\left(1-\frac{\cE''(p_0)}{v^2}\,Tx+\Or[(Tx)^2]\right).
\]
From the definition of $n_{\mathrm f,2}$ we immediately obtain
\begin{align*}
n_{\mathrm f,2}&=\frac T{\pi v}\int_0^{\infty}\frac{\diff x}{1+\e^{x}}
  -\frac{\cE''(p_0)T^2}{\pi v^3}\int_0^{\infty}\frac{x\,\diff x}{1+\e^{x}}
                 +\Or(T^3)\\
  &=\frac{\log 2}{\pi v}\,T-\frac{\pi\cE''(p_0)}{12v^3}\,T^2+\Or(T^3),
\end{align*}
and combining this result with Eq.~\eqref{nf1} we finally have
\begin{equation}
  \label{nfT0}
  n_{\mathrm f}=\frac{p_0}\pi-\frac{\pi\cE''(p_0)}{6v^3}\,T^2+\Or(T^3)\,.
\end{equation}
In particular, for the XX and $\su(1|1)$ HS chains the low temperature expansion~\eqref{nfT0}
reads
\begin{align*}
n_{\mathrm
  f,\mathrm{XX}}&=\frac2\pi\arcsin(\sqrt\la/2)-\frac{\pi(2-\la)}{6\la^{3/2}}\,T^2+\Or(T^3)\,,\\
  n_{\mathrm
  f,\mathrm{HS}}&=1-\frac1\pi\sqrt{\pi^2-2\la}+\frac{\pi T^2}{6(\pi^2-2\la)^{3/2}}+\Or(T^3)\,.
\end{align*}
The absence of a term linear in~$T$ in Eq.~\eqref{nfT0} is in agreement with the low temperature
behavior of~$n_{\mathrm f}$ apparent from Fig.~\ref{fig.nfplots}. It is also interesting to
observe that the sign of the leading correction to the $T=0$ value of~$n_{\mathrm f}$ is opposite
to that of the second derivative of~$\cE$ at the Fermi momentum~$p_0$. This behavior can be
understood by noting that the energy difference between adding a fermion with momentum $p_0+\De p$
(or $2\pi-p_0-\De p$) and removing a fermion with momentum $p_0-\De p$ (or $2\pi-p_0+\De p$),
with~$0<\De p\ll1$, is given by $\cE(p_0+\De p)+\cE(p_0-\De p)-2\cE(p_0)\simeq\cE''(p_0)\De p^2$.
Thus when $\cE''(p_0)<0$ the addition of a fermion is energetically more favorable than its
removal for momenta close to the Fermi momentum $p_0$ (or to $2\pi-p_0$), and consequently the
fermion density should increase at sufficiently low temperatures. For instance, for the elliptic
interaction~\eqref{hell} with~$\al\ge0$ finite $\cE''(p)$ is positive for $p$ less than a critical
momentum (which depends on~$\al$) and negative for larger momenta, while for the $\su(1|1)$ HS
chain (i.e., for~$\al=\infty$)~$\cE''(p)=-1$ is always negative. Again, these facts are consistent
with the behavior of~$n_{\mathrm f}$ observed in Fig.~\ref{fig.nfplots}.

\section{Summary and outlook}\label{sec.sumout}
In this paper we introduce a general class of $\su(1|1)$ supersymmetric spin chains with
long-range interactions generalizing the $\su(1|1)$ Haldane--Shastry and Inozemtsev (elliptic)
chains, which can be fermionized using the algebraic properties of the~$\su(1|1)$ permutation
operator. We exploit this fact to study the critical behavior of this class of models (with
nonzero chemical potential $\la$) in terms of their dispersion relation~$\cE(p)$. More precisely,
we show that they are gapless when the chemical potential lies on the interval~$[0,\cE(\pi)]$, and
that their ground state is a product state unless $\la$ belongs to this interval. We prove that
the models under study are actually critical when~$0<\la<\cE(\pi)$ by verifying that their low
energy excitations are linear in the excitation momentum, and that their free energy at low
temperature exhibits the characteristic quadratic behavior found in a $(1+1)$-dimensional CFT with
$c=1$~\cite{BCN86,Af86}. As further confirmation of this critical behavior, we find an exact
asymptotic formula for the von Neumann and R\'enyi entanglement entropies for the ground state,
showing that when~$\la$ belongs to the open interval~$(0,\cE(\pi))$ they both scale as~$\log L$
when the size~$L$ of the block of spins considered tends to infinity. Moreover, in both cases the
constant multiplying~$\log L$ is the same as for a $(1+1)$-dimensional CFT with central
charge~$c=1$~\cite{HLW94,CC04JSTAT,CC05}. Likewise, we show that the asymptotic behaviors of the
entanglement entropy and the zero-temperature fermion density as~$\la$ approaches the endpoints of
the critical interval $(0,\cE(\pi))$ are consistent with a quantum (continuous) phase transition.
We also analyze the fermion density at finite temperature for a particular class of models with
elliptic interactions, finding that its behavior is nontrivial (for instance, it is not always a
monotonic function of the temperature, and it can in fact present up to two extrema at finite
temperature) when~$\la$ belongs to the critical interval.

The results of this paper suggest several lines for future research. For one thing, they might
prove relevant for the~$\su(2)$ analogs of the models discussed in this paper, and most notably
the~spin~$1/2$ Inozemtsev and HS chains in the presence of a magnetic field. Indeed, it has been
analytically shown that the~$\su(1|1)$ HS chain with zero chemical potential~$\la$ is equivalent
in the thermodynamic limit to its~$\su(2)$ counterpart with zero magnetic field~\cite{FG14JSTAT}.
More recently, a numerical computation of the free energy of the~spin~$1/2$ elliptic chain with no
magnetic field suggests that this model and its~$\su(1|1)$ version with $\la=0$ studied in this
paper are also equivalent in the thermodynamic limit~\cite{Kl16}. If this equivalence could be
extended to the case of non-zero chemical potential (or magnetic field strength, for the~$\su(2)$
models), the results of this work could be used, for instance, to evaluate the ground state
entanglement entropy of the~spin~$1/2$ elliptic chain and its asymptotic limit when $L$ tends to
infinity.

Another line of research suggested by the present work is the study of the entanglement entropy of
the low-lying states of the $\su(1|1)$ supersymmetric model~\eqref{Hchain} when the chemical
potential lies in the critical interval~$(0,\cE(\pi))$. Indeed, it has been recently
shown~\cite{CIS11,ICS12} that in a $(1+1)$-dimensional CFT the quotient between the entanglement
entropy of an excited state created by acting on the vacuum with a primary field and that of the
ground state is a universal quantity, essentially determined by the conformal weights of the
field. Thus the computation of the entanglement entropy of the lowest excited states of the
model~\eqref{Hchain} when~$\la\in(0,\cE(\pi))$, which can be constructed from the equivalent
fermionic model~\eqref{Hf}, could shed some light on its underlying CFT.

\section*{Acknowledgments}
This work was partially supported by Spain's MINECO under grant no.~FIS2015-63966, and by the
Universidad Complutense de Madrid and Banco Santander under grant no.~GR3/14-910556. PT has been
partly supported by the ICMAT Severo Ochoa project SEV-2015-0554 (MINECO). JAC would also like to
thank the Madrid township and the ``Residencia de Estudiantes'' for their financial support.

% \bibliographystyle{apsrev4-1}
% \bibliography{cmprefs}


\end{document}